\documentclass[12pt
]{iopart}
\usepackage{graphicx,iopams,bm}
\newcommand{\be}{\begin{equation}}
\newcommand{\ee}{\end{equation}}
\newcommand{\bea}{\begin{eqnarray}}
\newcommand{\eea}{\end{eqnarray}}

\newcommand{\nn}{\nonumber}
\def\rf#1{(\ref{#1})}

\newcommand{\sgn}{ \mathrm{sgn}}
\def\udot{\dot{u}}
\newcommand{\sfrac}[2]{{\textstyle{#1\over#2}}}

\def\Th{\Theta}
\def\sig{\sigma}
\def\om{\omega}
\def\lgl{\langle}
\def\rgl{\rangle}
\def\de#1/de#2{\frac{\partial {#1}}{\partial {#2}}}
\begin{document}

\title{The dynamics of  Bianchi I universes in $R^n$ cosmologies with torsion.}

\author{ Sante Carloni$^{1}$\footnote{E-mail: sante.carloni@esa.int}\, Stefano Vignolo$^{2}$\footnote{E-mail: vignolo@diptem.unige.it} \ and
Luca Fabbri$^{2,3}$\footnote{E-mail: fabbri@diptem.unige.it}
}
\address{ $^{1}$ESA-Advanced Concept team, European Space Research Technology Center (ESTEC)\\
Keplerlaan 1, Postbus 299, 2200 AG Noordwijk The Netherlands.}
\address{ $^{2}$DIME Sez. Metodi e Modelli Matematici, Universit\`{a} di Genova\\
Piazzale Kennedy, Pad. D - 16129 Genova (Italia)
}
\address{$^{3}$INFN \& Dipartimento di Fisica, Universit\`{a} di Bologna\\
Via Irnerio 46, - 40126 Bologna (Italia)}

\begin{abstract}
We analyze the phase space of Bianchi I cosmologies filled by a spin fluid in the framework of $f(R)$-gravity with torsion using a combination of the dynamical systems approach and the 1+3 covariant formalism.  In the simple case of $f(R)=R^n$ our results allow a quantification of the role of torsion and the spin of the cosmic fluid in the evolution of the cosmology.  While torsion is able to modify the cosmological dynamics with respect to the purely metric case, the spin has little influence on the cosmology. We argue that this is due to the different symmetries of the tensor characterizing the anisotropies and the spin tensor. The cosmological model we analyzed presents isotropization for a wide set of initial conditions and values of the parameters and allows for two types of exotic bounce solutions.
\end{abstract}
\date{\today}
\submitto{\CQG}
\maketitle

\tolerance=5000

%%%%%%%%%%%%%%%%%%%%%%%
\section{Introduction}
%%%%%%%%%%%%%%%%%%%%%%%
In the quest for the understanding of the (now not so) new evidence about the increase in the expansion rate of the Universe dubbed cosmic acceleration, modifications of the structure of Einsteinian gravitation have played a major (and highly debated) role \cite{Sotiriou2010,Odintsov2003}.  Although it is still not clear if this unexpected behavior of the Universe should or could be ascribed to new aspects of the gravitational interaction, so far a great number of modified theories of gravitation have been proposed. %These models present a series of interesting feature such as the existence of solutions that can represent cosmic acceleration, but they are also are significantly more complicated to analyze than General Relativity.  
The most studied classes of such extensions are the ones that are often called "extended theories of gravity". These theories are in some sense ``minimal'' modifications of GR and are designed to be fully metric--affine i.e. such that they are completely determined by a metric and a connection \cite{Capozziello2008}.

In this work we will focus on one type of such extensions of General Relativity (GR), which takes the name of $f(R)$-gravity \cite{revnostra}  and it is characterized by an action which is non-linear in the Ricci scalar. These theories emerged for the first time in quantum field theory in curved spacetime as an effect of the renormalization of the matter stress energy tensor \cite{B&D} and were successively recovered as low energy limit of fundamental theories  like string theory \cite{Nojiri2003}. Many aspects of the cosmology of $f(R)$-gravity have been investigated with a variety of techniques, but most of the studies were focused on the homogeneous and isotropic case. Even in this simple case the differences between the cosmology of Relativity and the ones of these models can be quite dramatic. It is known for example that in the some simple models of this type perturbations can grow even in a state of accelerated expansion \cite{Carloni:2007yv}.

All of these extended models of relativistic gravitation are usually designed considering matter as a purely classical system, neglecting quantum properties like, for example, the spin. Nonetheless the spin is an important property of the fermions, which are the most common type of fundamental constituents of standard matter. It is natural then to attempt the introduction of this feature in  the mathematical structure of relativistic gravitation.  In the context of GR it has been proven that one way to achieve this goal   is to include  an additional geometric property of spacetime: torsion \cite{Hehl,Obukhov,Hehl-Heyde-Kerlick}.  Geometrically torsion is defined as a property of the spacetime that prevents infinitesimal parallelograms to be closed. This idea translates directly in the existence of an antisymmetric part of the connection and it is the cornerstone of  the Einstein-Cartan-Sciama-Kibble (ECSK) theory. Since the choice to have only a completely symmetric connection was an assumption made in the first years of GR, we can think about ECSK theory as a generalization rather than a modification of Einstein gravitation. The presence of torsion beside curvature, allows both the spin  and the energy of matter fields to be coupled with gravitation and opens naturally the way to the construction of an Einstein-Dirac theory.

The inclusion of torsion is therefore justified by the presence of the spin, which is itself justified since we know about the existence of fermions. Within fermionic field equations in general, and for the Dirac field in particular, the coupling spin--torsion gives rise to fermionic self-interactions \cite{n-j--l}.
In the framework of ECSK theory, however, since the gravitational Lagrangian is the usual Hilbert-Einstein one (with torsion) containing only the gravitational Newton constant, the torsionally-induced effects would be manifest  only at the Planck scale.  
One can, of course, consider more general torsional completions of the Hilbert-Einstein action. One possibility is to introduce squared torsional contributions in the action so that the torsional terms will have their own coupling constant \cite{Fabbri:2011kq,Fabbri:2012yg,VFS}. In these theories  torsion is  present explicitly as an independent field and it has its own coupling constant which can be much larger than the Newton one, giving rise to torsional effects at much larger scales.

In analogy with GR, one can also think to include semi-classical corrections in the ECSK theory so to obtain an action non linear in the scalar curvature. If we allow such an extension even more dramatic effects will emerge. In particular, the torsional contributions would acquire a coupling constant that does not only need to be fixed at scales that may be the nuclear one but it may also display a running behavior \cite{Capozziello:2012gw}.

Up to now most of the analysis of these non linear ECSK theories has been focused on homogeneous and isotropic cosmologies, but the symmetries of these spacetimes greatly limit the effect of the additional torsional contributions. For this reason it is interesting to analyze these theories in more complicated spacetimes. One could think therefore to investigate the next level of generalization: the Bianchi spacetimes. Bianchi spacetimes are characterized by homogeneity, but present different degrees of anisotropies. They are classified according to the symmetry groups of the spacetime in nine different classes that include, for example, the Taub metric \cite{Taub}. However, since the anisotropies are characterized by a symmetric tensor and the torsion/spin contributions are antisymmetric by definition, it is unlikely that their interaction will be determinant in Bianchi cosmologies. The one exception is constituted by the Bianchi I universes in which both the anisotropies and the spin terms can be represented by scalar functions and can therefore influence each other. Hence Bianchi I universes offer a relatively simple testing ground for the analysis of the effect of spin in spacetimes with torsion.

In this paper we will analyze the dynamics of Bianchi I universes in the framework of $f(R)$-gravity with torsion, in which the gravitational Lagrangian is a non linear function of the scalar curvature in presence of torsion. The aim is to estimate the influence that these non linear terms have on the behavior of the cosmology and in particular on the development of the anisotropies. We will perform this analysis by means of the Dynamical Systems Approach (DSA). This approach  was proposed for the first time in its final form by Wainwright and Ellis and had a key role in the analysis of the dynamics of Bianchi universes \cite{Dynamical} . In the last few years this technique has been adapted to the investigation of the cosmologies of modified gravity revealing a series of interesting features \cite{Carloni:2004kp,Leach:2006br,Carloni:2006mr,Carloni:2007eu,Abdelwahab:2007jp,Carloni:2007br,Roy:2011za,Bonanno:2011yx,Goliath:1998na}. In the present paper the DSA will be combined with the {\it 1+3 covarian approach}. This formalism was first proposed in \cite{Ehlers,Kundt,carge73,Cargese} and it has been used to analyze a wide variety of different problems including the dynamics of linear cosmological perturbations \cite{EB,EBH,BDE,DBE,BED,Challinor:1998aa} and the investigation of the features of modifications of the gravitational interaction \cite{Carloni:2007yv,Carloni:2006gy,Carloni:2006fs,Carloni:2010tv,Abebe:2011ry}.

The paper will be organized as follows. In Section II we will review  the general $f(R)$-gravity theory with torsion coupled to spin fluids. In Section III we will write the key cosmological  equations in the 1+3 covariant approach on which the dynamical systems approach will based.  Section V will be dedicated to the phase space analysis of the cosmological equations. Finally, Section VI will be dedicated to the conclusions.

Unless otherwise specified, natural units ($\hbar=c=k_{B}=8\pi G=1$) will be used throughout this paper, Latin indices run from 0 to 3. The symbol $\nabla$ represents the  covariant derivative and $\partial$ corresponds to partial differentiation in the Riemann-Cartan space-time\cite{Hehl,Obukhov,Hehl-Heyde-Kerlick}. We use the $-,+,+,+$ signature and the Riemann tensor is defined by
\begin{equation}
\mathcal{R}^{d}_{\;\;cab}
=\partial_a\Gamma_{bc}^{\;\;\;d} - \partial_b\Gamma_{ac}^{\;\;\;d} +
\Gamma_{ap}^{\;\;\;d}\Gamma_{bc}^{\;\;\;p}-\Gamma_{bp}^{\;\;\;h}\Gamma_{ac}^{\;\;\;p}\;,
\end{equation}
where the $\Gamma_{ij}^{\;\;\;h}$ are the coefficients of the linear dynamical connection, defined by
\begin{equation}
\nabla_{\partial_a}\partial_b = \Gamma_{ab}^{\;\;\;c}\,\de /de{x^c}\,.
\end{equation}
The Ricci tensor is obtained by contracting the {\em first} and the {\em third} indices via the metric $g_{ab}$
\begin{equation}\label{Ricci}
\mathcal{R}_{ab}=\mathcal{R}^{c}_{\phantom{c}acb}\,.
\end{equation}
The purely metric Riemann and Ricci tensors will be defined with the same way using the metric connection $\tilde{\Gamma}_{ij}^{\;\;\;h}$ and will be indicated by $R_{a}^{\;bcd..},R_{ab}, etc.$. In general all the metric quantities will be indicated with a tilda.
Finally, the symmetrization and the antisymmetrization  over the indexes of a tensor are defined as 
\begin{equation}
T_{(a b)}= \frac{1}{2}\left(T_{a b}+T_{b a}\right)\;,\qquad T_{[a b]}= \frac{1}{2}\left(T_{a b}-T_{b a}\right)\,.
\end{equation}
%Finally the Hilbert--Einstein action in the presence of matter is given by
%\begin{equation}
%{\cal A}=\frac12\int d^4x \sqrt{-g}\left[R+ 2{\cal L}_m \right]\;.
%\end{equation}

%%%%%%%%%%%%%%%%%%%%%%%%%%%%%%%%%%%%%%%%%%%%%%%%%%%%%%%%%%%%%%%%%%%%%%%%%%%%%%%%%%%%%%%%%%%%%%%%%%
\section{$f(\mathcal{R})$-gravity with torsion coupled to spin fluids}
%%%%%%%%%%%%%%%%%%%%%%%%%%%%%%%%%%%%%
The field equations of $f(\mathcal{R})$-gravity with torsion are \cite{CCSV1,CV4}
\begin{equation}
\label{2.1a}
f'\/(\mathcal{R})\mathcal{R}_{ab} -\frac{1}{2}f\/(\mathcal{R})g_{ab}=M_{ab},
\end{equation}
\begin{equation}
\label{2.1b}
T_{ab}^{\;\;\;c}
=\frac{1}{f'(\mathcal{R})}
\left[\frac{1}{2}\left(\de{f'(\mathcal{R})}/de{x^{p}}+S_{pq}^{\;\;\;q}\right)
\left(\delta^{p}_{b}\delta^{c}_{a}-\delta^{p}_{a}\delta^{c}_{b}\right)
+S_{ab}^{\;\;\;c}\right].
\end{equation}
In eqs. (\ref{2.1a}-\ref{2.1b}) $\mathcal{R}_{ab}$ and $T_{ab}^{\;\;\;c}$ are the Ricci and Torsion tensors associated with the dynamical gravitational variables $(g,\Gamma)$, where $g$ is a metric tensor and $\Gamma$ is a metric--compatible linear connection; $M_{ab}$ and $S_{ab}^{\;\;\;c}$ are the energy-momentum and spin density tensors of the matter fields. 
\\
In general, the trace of equations \rf{2.1a}
\begin{equation}\label{2.1bis}
f'(\mathcal{R})\mathcal{R} -2f(\mathcal{R})=M,
\end{equation}
is supposed to give rise to an invertible relation between the Ricci scalar curvature $\mathcal{R}$ and the trace $M$ of the energy-momentum tensor. Also, it is assumed that $f(\mathcal{R})\not = k\mathcal{R}^2$ (we notice that the case $f(\mathcal{R})=k\mathcal{R}^2$ is only compatible with the condition $M=0$). Under the stated conditions, from equation \rf{2.1bis} it is possible to express the Ricci scalar curvature $\mathcal{R}$ as a suitable function of $M$, namely
\begin{equation}\label{2.1tris}
\mathcal{R}=\mathcal{R}(M).
\end{equation}
Through the Bianchi identities \cite{FV1}, or also from the invariance under diffeomorphisms and Lorentz transformations \cite{Poplawski_conservation_law}, it is possible to obtain the conservation laws of the theory 
\begin{equation}
\label{2.2a}
\nabla_{a}M^{ab}+T_{a}M^{ab}-M_{pa}T^{bpa}-\frac{1}{2}S_{sta}\mathcal{R}^{stab}=0,
\end{equation}
\begin{equation}
\label{2.2b}
\nabla_{c}S^{abc}+T_{c}S^{abc}+M^{ab}-M^{ba}=0.
\end{equation}
%In eqs. (\ref{2.2a}-\ref{2.2b}) $\nabla_a$ denotes covariant derivative with respect to the dynamical connection $\Gamma$, while $\mathcal{R}^{stij}$ indicates the curvature tensor of $\Gamma$; indices are raised and lowered by the metric $g$.
%\\
Moreover, we recall that any $g$ metric--compatible connection $\Gamma$ can be expressed as the sum
\begin{equation}\label{2.2bis}
\Gamma_{ab}^{\;\;\;c} = \tilde{\Gamma}_{ab}^{\;\;\;c} - K_{ab}^{\;\;\;c},
\end{equation}
where %$\tilde{\Gamma}_{ab}^{\;\;\;c}$ are the coefficients of the Levi--Civita connection induced by the metric $g_{ab}$ and 
$K_{ab}^{\;\;\;c}$ are the components of the contortion tensor \cite{Hehl}
\begin{equation}\label{2.3}
K_{ab}^{\;\;\;c} = \frac{1}{2}\/\left( - T_{ab}^{\;\;\;c} + T_{b\;\;\;a}^{\;\;c} - T^c_{\;\;ab}\right).
\end{equation}
Making use of eqs. \rf{2.2bis} and \rf{2.3}, from the field equations \rf{2.2b} we obtain then the following representations
\begin{equation}\label{2.4a}
K_{ab}^{\;\;\;c}= \hat{K}_{ab}^{\;\;\;c} + \hat{S}_{ab}^{\;\;\;c},
\end{equation}
\begin{equation}\label{2.4b}
\hat{S}_{ab}^{\;\;\;c}:=-\frac{1}{2f'(\mathcal{R}(M))}\/\left(  S_{ab}^{\;\;\;c} - S_{b\;\;\;a}^{\;\;c} + S^c_{\;\;ab}\right),
\end{equation}
\begin{equation}\label{2.4c}
\hat{K}_{ab}^{\;\;\;c} := -\hat{T}_b\delta^c_a + \hat{T}_pg^{ph}g_{ab},
\end{equation}
\begin{equation}\label{2.4d}
\hat{T}_b:=\frac{1}{2f'(\mathcal{R}(M))}\/\left( \de{f'}/de{x^b} + S^{\;\;\;p}_{b p} \right),
\end{equation}
which, together with equation \rf{2.2bis}, enable us to decompose the field equations (\ref{2.1a}-\ref{2.1b}) in purely metric and torsional terms.
\\
In the present paper, we consider $f(\mathcal{R})$-gravity coupled to a Weyssenhoff spin fluid; the latter is characterized by an energy-momentum tensor of the form
\begin{equation}\label{2.5.1a}
M^{ab}= u^aP^b + p\left( u^au^b + g^{ab}\right),
\end{equation}
and a spin density tensor given by
\begin{equation}\label{2.5.1b}
S_{ab}^{\;\;\;c}=S_{ab}u^c,
\end{equation}
where $u^a\/$ ($u^au_a=-1\/$) and $P^a$ denote respectively the $4$-velocity and the $4$-vector density of energy-momentum, while $S_{ab}\/$ the spin density of the fluid (see, for example, \cite{Obukhov,Hehl-Heyde-Kerlick} and references therein). The $4$-velocity and the spin density are assumed to satisfy the convective condition
\begin{equation}\label{2.5.2}
S_{ab}u^b =0.
\end{equation}
The relations \rf{2.5.2}, together with equation \rf{2.4b}, imply the identities 
\begin{equation}\label{2.5.2bis}
\hat{S}_a^{\;\;ac}=-\hat{S}_a^{\;\;ca}=0.
\end{equation}
Making use of eqs. \rf{2.2bis}, \rf{2.3}, (\ref{2.4a}-\ref{2.4d}), \rf{2.5.1b}, \rf{2.5.2}, \rf{2.5.2bis} and denoting by $\tilde\nabla$ the Levi--Civita covariant derivative, we can decompose the contracted curvature and the scalar curvature respectively as \cite{VF2}
\begin{eqnarray}\label{2.5.3a}
&&\nn \mathcal{R}_{ab} = R_{ab} - 2\tilde{\nabla}_{b}\hat{T}_a - \tilde{\nabla}_c\hat{T}^cg_{ab} + 2\hat{T}_a\hat{T}_b - 2\hat{T}_c\hat{T}^cg_{ab} - \frac{1}{f'}\hat{T}_cS^c_{\;\;b}u_a+ \\
&&~~~~~~~~~- \frac{1}{2f'}\tilde{\nabla}_c\/\left(- S_{ba}u^c + S_a^{\;\;c}u_b - S^c_{\;\;b}u_a\right) + \frac{1}{4(f')^2}S^{pq}S_{pq}u_au_b,
\end{eqnarray}
and
\begin{equation}\label{2.5.3b}
\mathcal{R} = R - 6\tilde{\nabla}_{a}\hat{T}^a - 6\hat{T}_a\hat{T}^a - \frac{1}{4(f')^2}S^{pq}S_{pq},
\end{equation}
where now, due to the convective condition, we have $\hat{T}_a =\frac{1}{2f'}\de{f'}/de{x^a}\/$. Substituting expressions (\ref{2.5.3a}-\ref{2.5.3b}) into equations (\ref{2.1a}-\ref{2.1b}), we get Einstein-like equations of the form \cite{VF2}
\begin{eqnarray}\label{2.5.4}
&& \nn \fl R_{ab} -\frac{1}{2}Rg_{ab}= \\  &&  \nn\fl \frac{1}{f'({\mathcal R)}}M_{ab} + \frac{1}{f'({\mathcal R})}\left[\left(g_a^c g_b^d-g^{cd}g_{ab}\right)\tilde{\nabla}_c\tilde{\nabla}_{d}f'({\mathcal R})-\frac{3}{2f'({\mathcal R})}\left(g_a^c g_b^d-\frac{1}{2}g^{cd}g_{ab}\right)\times\right. \\&& 
\nn \fl \left. \tilde{\nabla}_c f'({\mathcal R})\tilde{\nabla}_{d}f'({\mathcal R})-\frac{1}{4}\left(M+{\mathcal R}f'({\mathcal R})\right)g_{ab} \right]+ \frac{1}{f'({\mathcal R})}\hat{T}_cS^c_{\;\;b}u_a +
 \\&& \fl + \frac{1}{2 f'({\mathcal R})}\tilde{\nabla}_c\/\left(- S_{ba}u^c + S_a^{\;\;c}u_b - S^c_{\;\;b}u_a\right)
- \frac{1}{4f'({\mathcal R})^2}S^{pq}S_{pq}u_au_b - \frac{1}{8\varphi^2}S^{pq}S_{pq}g_{ab}.
\end{eqnarray}
The anti-symmetrized part of the Einstein-like equations \rf{2.5.4} is equivalent to the conservation laws for the spin \rf{2.2b} 
\begin{equation}\label{2.5.9}
\tilde{\nabla}_c\/\left(S_{ab}u^c \right) + \hat{T}_cS^c_{\;\;b}u_a - \hat{T}_cS^c_{\;\;a}u_b + M_{ab} - M_{ba} =0.
\end{equation}
Saturating equations \rf{2.5.9} with $u^a$, we get the explicit expression for the $4$-vector density of energy-momentum
\begin{equation}\label{2.5.10}
P_b = \rho\,u_b - \hat{T}_c\/S^c_{\;\;b} + \tilde{\nabla}_c\/\left(S_{ab}u^c \right)u^a,
\end{equation}
where $\rho := - u^aP_a$. Inserting expressions \rf{2.5.10} into \rf{2.5.1a}, we obtain the explicit form of the energy-momentum tensor
\begin{equation}\label{2.5.11}
M_{ab} = (\rho + p)\,u_au_b + p\,g_{ab} - u_a\hat{T}_c\/S^c_{\;\;b} + u_a\tilde{\nabla}_c\/\left(S_{pb}u^c \right)u^p.
\end{equation}
The symmetric part of the Einstein-like equations \rf{2.5.4}  instead can be written as:
\begin{eqnarray}
&&  R_{ab}-\frac{1}{2}R g_{ab}=
M_{ab}^{tot}=M^{\mathcal R}_{ab}+{\bar M}^m_{ab}+M^S_{ab}, \label{FieldEqTorsR}\\
&& \nn  M_{ab}^{{\mathcal R}}=\frac{1}{f'({\mathcal R})}\left[\left(g_a^c g_b^d-g^{cd}g_{ab}\right)\tilde{\nabla}_c\tilde{\nabla}_{d}f'({\mathcal R})-
\frac{3}{2f'({\mathcal R})}\left(g_a^c g_b^d-\frac{1}{2}g^{cd}g_{ab}\right)\times  \right. \\&&~~~~~~~~~  \left. \tilde{\nabla}_c f'({\mathcal R})\tilde{\nabla}_{d}f'({\mathcal R})-
\frac{{\mathcal R}f'(\mathcal{R})}{4}g_{ab} \right],\label{SigEff}\\
&&  \bar{M}^{m}_{ab}=\frac{1}{f'({\mathcal R})}\left(M^m_{ab}-\frac{1}{4}g_{ab}M^m\right),\label{Matter}\\
&&  \nn M^S_{ab}=-\frac{1}{f'({\mathcal R})}\tilde{\nabla}_{c}\left(S^c_{(a}u_{b)}\right)+
\frac{1}{2f'({\mathcal R})^2} \tilde{\nabla}_{d}f'({\mathcal R})S^{d}_{(a}u_{b)}+\\&&~~~~~~~~~ -
\frac{1}{4f'({\mathcal R})^2}S_{cd}S^{cd}\left(\frac{1}{2}g_{ab}+u_au_b\right),
\end{eqnarray}
which will be our starting point.  We will define for later convenience ${\mathcal S}=\frac{1}{2}S^{ab}S_{ab}$.

%%%%%%%%%%%%%%%%%%%%%%%%%%%%%%%%%%%%%%%%%%%%%%%%%%
\section{1+3 $f(R)$-Gravity with Torsion and Spin}
%%%%%%%%%%%%%%%%%%%%%%%%%%%%%%%%%%%%%%%%%%%%%%%%%%
In this section we will rewrite the equations of the previous section using the main tool of this paper: the {\it 1+3 covariant approach}.  Let us start with a brief summary of the method. Such summary will not by any means be complete and we refer the reader to \cite{Ehlers,Kundt,carge73,Cargese} and references therein for more detailed information.  

For any given fluid 4-velocity vector field $u^{a}$, the projection tensor
$h_{ab}=g_{ab}+u_{a}u_{b}$ projects into the instantaneous
rest-space of a comoving observer \cite{carge73,Cargese}. The first covariant derivative
can be decomposed as 
\begin{equation}
 \nabla_{a}u_{b} = -\,u_a\,\udot_b +
D_{a}u_{b} = -\,u_a\,\udot_b + {\sfrac{1}{3}}\,\Th\,h_{ab} +
\sigma_{ab} + \om_{ab} \ , \label{eq:kin}
\end{equation}
where  $\dot{X}=u^c\nabla_c X$, $D_{a}X_{bc}=h^{d}_{a} h^{e}_{a} h^{f}_{a} ... \nabla_{d} X_{e f...}$, $\sig_{ab}$ is the symmetric shear tensor ($\sig_{ab} =
\sig_{(ab)}$, $\sig_{ab}\,u^b = 0$, $\sig^a{}_{a}= 0$), $\om_{ab}$
is the vorticity tensor ($\om_{ab} = \om_{[ab]}$, $\om_{ab}\,u^b =
0$) and $\udot_a$ is the acceleration vector
($\udot_a=u^b\nabla_{b}u_{a}$). Finally, $\Theta$ is the volume expansion
($\Theta =\nabla^au_a$) which defines a length scale $a$ along the
flow lines via the standard relation $\Theta =\sfrac{3\dot{a}}{a}$.

A general matter energy-momentum tensor $M_{ab}$ can always be uniquely decomposed
with respect to $u^a$ in the form
\begin{equation}
M_{ab} = \mu\,u_a\,u_b + q_a\,u_b + u_a\,q_b + p\,h_{ab} +
\pi_{ab} \label{eq:stress}  \ , \nonumber
\end{equation}
where $\mu$ is the relativistic energy density, $p$ the isotropic
pressure, $q^{a}$ the energy flux ($q_a\,u^a = 0$) and $\pi_{ab}$
the trace-free anisotropic pressure ($\pi^a{}_a = 0$, $\pi_{ab} =
\pi_{(ab)}$, $\pi_{ab}\,u^b = 0 $), all relative to $u^a$.

One can therefore derive the general propagation and constraint equations for the variables given above. These equations are called {\it1+3 covariant equations} and their full set can be found in \cite{Cargese}. 

Here, we consider Bianchi I spacetimes. They are characterized by homogeneous hypersurfaces
have isotropic 3-curvature $\tilde{R} _{ab}=\sfrac{1}{3}\tilde{R} h_{ab}=0$. In addition, spatial homogeneity implies that the spatial gradients will
vanish and that $\udot_a=0=\omega$. For the same reason $q_a$ is identically zero. Thus the characterizing kinematical 1+3 equations 
for this cosmological model are the Raychaudhuri equation
\begin{equation}\label{Ray:f(R)}
 \dot{\Th} + \sfrac{1}{3}\,\Th^{2} + 2\,\sig^{2} 
+\frac{1}{2}\left(\mu+3p\right)
=0,
\end{equation}
the shear equation
\begin{equation}\label{Shear:f(R)}
\dot{\sigma}_{\langle {{a}}{{b}} \rangle }
+{\textstyle\frac{2}{3}}\Theta\sigma_{{{a}}{{b}}}
+\sigma_{{c}\langle {{a}}}\sigma_{{{b}}\rangle }{}^{c}
 -\frac{1}{2}\pi_{{{a}}{{b}}}=0,
\end{equation}
the definition of the 3-Ricci tensor
\begin{eqnarray}\label{3Ricciab}
0=\tilde{R} _{ab}&=& - \dot{\sig}_{\lgl
ab\rgl}-\Th\,\sig_{ab} +\pi_{ab} - \sfrac{1}{3}h_{ab}\left( 2\sigma^2-\sfrac{2}{3}\Th^2+2\mu\right),
\end{eqnarray}
and its trace
\begin{equation}\label{3Ricci}
0=\tilde{R} = 2\sigma^2-\sfrac{2}{3}\Th^2+2\mu.
\end{equation}
where  $\sigma_{ab}\sigma^{ab}=2\sigma^2$ and $T_{\lgl ab\rgl}=T_{cd}\left(h^{c}_{(a}h^{d}_{b)}-\frac{1}{3}h^{cd}h_{ab}\right)$.

For our purposes it will be useful to write \rf{3Ricciab} and \rf{3Ricci} as an equation for the shear propagation and a constraint between the shear, the expansion and the matter-energy density:
\begin{eqnarray}
&&\dot{\sig}_{\lgl a b \rgl}+\Th\,\sig_{ab} -\pi_{ab}=0,\label{Shear-ab} \\
&& \sfrac{2}{3}\Th^2=2\sigma^2+2\mu.\label{Fried}
\end{eqnarray}
The first of the above equations can be used to write a dynamical equation for $\sigma$:
\begin{equation} \label{}
\dot{\sig}=-\Th \sig+ \frac{1}{2} \frac{\sigma_{ab}\pi^{ab}}{\sigma},
\end{equation}
and will be employed in place of  \rf{Shear:f(R)}.
The conservation equations $\nabla^bM_{ab}=0$, can be split with
respect to $u^a$ and $h_{ab}$:
\begin{eqnarray}
\dot{\mu}  +\Th(\mu+p) +
(\sig^{a}\!_{b}\pi^{b}\!_{a})=0, \label{eq:cons1}\\
 D^{a}p + D_{b}\pi^{ab} =0. \label{eq:cons2}
\end{eqnarray}
The set of equations \rf{Ray:f(R)}, \rf{Shear-ab}, \rf{Fried}, \rf{eq:cons1} form a closed system and are sufficient to describe completely the evolution of Bianchi I universes.

Let us now specialize these equations in the case of $f(R)$-gravity with torsion and spin. Like in the purely metric case \cite{revnostra,Carloni:2004kp} this is done recasting the gravitational field equations in a form that resembles standard Einstein gravity plus matter and an effective fluid encoding the properties of the higher order part of the gravitational interaction. For this effective fluid  the effective thermodynamics is characterized by 
\begin{eqnarray}
&& \fl \mu^{tot}=M^{tot}_{ab}u^{a}u^{b}=-\frac{f''({\mathcal R})}{f'({\mathcal{R}})}\dot{{\mathcal R}}\left[\Th+\frac{3}{4}\frac{f''({\mathcal R})\dot{{\mathcal R}}}{f'({\mathcal{R}})}\right]+\frac{1}{4}{\mathcal{R}}+\frac{3(\mu^m+p^m)}{4f'({\mathcal{R}})}-\frac{ {\mathcal S}^2}{4f'({\mathcal{R}})^2},\\
&& \nn \fl p^{tot}=M^{tot}_{ab}h^{ab}= \frac{f''({\mathcal R})}{f'({\mathcal{R}})}\dot{{\mathcal R}}\left[\frac{2}{3}\Th-\frac{3}{4}\frac{f''({\mathcal R})\dot{{\mathcal R}}}{f'({\mathcal{R}})} \right]+\frac{f''({\mathcal R})}{f'({\mathcal{R}})}\ddot{{\mathcal R}} +\frac{f'''({\mathcal R})}{f'({\mathcal{R}})}\dot{{\mathcal R}}^2-\frac{1}{4}{\mathcal{R}}\\ &&\quad +\frac{1}{4f'({\mathcal{R}})}(\mu^m+p^m)-\frac{{\mathcal S}^2}{4f'({\mathcal{R}})^2},\\
&&\fl q_{a}^{tot}=M^{tot}_{cd}h_{a}^{c}u^{d}=0,\\
&&\fl \pi_{ab}^{tot}=M^{tot}_{\langle cd\rangle}=-\sigma_{ab}\frac{f''({\mathcal R})\dot{{\mathcal R}}}{f'({\mathcal{R}})}.
\end{eqnarray}
Using these definitions in  the \rf{Ray:f(R)}, \rf{Shear-ab}, \rf{Fried}, \rf{eq:cons1} we obtain
\begin{eqnarray}
&& \fl\nn\dot{\Th} + \sfrac{1}{3}\,\Th^{2} + 2\,\sig^{2} +\frac{1}{2}\left\{\frac{f''(\mathcal{R
   })}{f'(\mathcal{R})} \left(3 \ddot{\mathcal{R}}+\Theta \dot{\mathcal{R}} \right)+3\left(\frac{ f^{(3)}(\mathcal{R})}{f'(\mathcal{R})}-\frac{ f''(\mathcal{R})^2}{f'(\mathcal{R})^2}\right) \dot{\mathcal{R}}^2+\right.\\
&&\fl~\qquad~\qquad~\qquad~\qquad~\qquad~~~\left.-\frac{\mathcal{R}}{2}+\frac{3}{2}\frac{\mu}{f'(\mathcal{R})}  \left(w+1\right)\right\}-\frac{{\mathcal S}^2}{2f'({\mathcal{R}})^2}=0, \label{EqBfR1}\\
&&\fl \dot{\sig}=-\Th \sig-\sig\frac{f''({\mathcal R})}{f'({\mathcal{R}})}\dot{{\mathcal R}}, \\
&& \fl \sigma^2-\sfrac{1}{3}\Th^2+\left\{\frac{1}{4}{\mathcal{R}}-\frac{f''({\mathcal R})}{f'({\mathcal{R}})}\dot{{\mathcal R}}\left[\Th+\frac{3}{4}\frac{f''({\mathcal R})\dot{{\mathcal R}}}{f'({\mathcal{R}})}\right]+\frac{3\mu^m(1+w)}{4f'({\mathcal{R}})}-\frac{{\mathcal S}^2 }{4f'({\mathcal{R}})^2}\right\}=0, \\
&& \fl \dot{\mu}^m +\Th\mu^m(1+w) =0,\\
&& \fl \dot{\mathcal S}=-\Theta{\mathcal S} \label{EqBfR5},
\end{eqnarray}
where the last two equations are a direct consequence of the (\ref{2.2a}-\ref{2.2b}) \cite{VF2}.

Because of the constraint \rf{2.1bis} these equations can be further simplified, but such simplification can be only obtained choosing a specific form for the function $f$. In the following we will analyze the case $f={\mathcal R}^n$.

%%%%%%%%%%%%%%%%%%%%%%%%%%%%%%%%%%%%%%%%%%%%%%%%%%
\section{ ${\mathcal R}^n$- Gravity with Torsion}
%%%%%%%%%%%%%%%%%%%%%%%%%%%%%%%%%%%%%%%%%%%%%%%%%%
We will now consider the simplest of the  possible forms of $f$ i.e. a general power of $\mathcal{R}$.  This model is simple enough to permit simple calculations and, at the same time preserves most of the most interesting features of the more general models. It will help us to gain an insight in the effect that torsion and spin have on the properties of the cosmic evolution. 

If $f({\mathcal R})=\alpha {\mathcal R}^n$ then using the \rf{2.1bis} we have
\begin{equation} \label{RofMuRn}
{\mathcal R}=\left(\frac{(3 w-1) \mu}{\alpha  (n-2)}\right)^{\frac{1}{n}},
\end{equation}
and the (\ref{EqBfR1}-\ref{EqBfR5}) can be written as
\begin{eqnarray}
&&\fl \dot{\Theta}=-\frac{1 }{3}\Theta ^2-\frac{ (n-2)^{-\frac{1}{n}} \alpha ^{-\frac{1}{n}}(-2 n+3 w+3) (3w-1)^{\frac{1}{n}}}{(3 w-1) (3 (n-1) w+n-3)}\mu ^{\frac{1}{n}} + \frac{4n }{(3 (n-1) w+n-3)}\sigma ^2 +\nn\\&&~~~~~~\fl -\frac{(n-2)^{2-\frac{2}{n}} \mathcal{S}^2 \alpha ^{-2/n} (3 w-1)^{\frac{2}{n}-2} \mu ^{\frac{2}{n}-2}}{n (3 (n-1) w+n-3)},\label{RayRnS}\\
&& \fl \dot{\sigma}=\frac{ (n w-w-1)}{n}\Theta  \sigma,\label{ShearRnS}\\
&& \fl \nn 2 \sigma ^2-\frac{ (3 (n-1)w+n-3)^2}{6 n^2}\Theta ^2+\frac{(n-2)^{-\frac{1}{n}} \alpha ^{-\frac{1}{n}}  (3 (n-1) w+n-3) (3 w-1)^{\frac{1}{n}}}{n(3 w-1)}\mu^{\frac{1}{n}}+\\&&~~~~~\fl -\frac{ (n-2)^{2-\frac{2}{n}}\alpha ^{-2/n} (3 w-1)^{\frac{2}{n}-2} \mu^{\frac{2}{n}-2}}{2 n^2} {\mathcal S}^2 =0,\label{HamiCosntr}\\
&& \fl \dot{\mu}^m+\Th\mu^m(1+w) =0,\label{MatConsS}\\
&& \fl \dot{\mathcal S}=- \Theta {\mathcal S},\label{SpinEqRn}
\end{eqnarray}
At this point we are ready to analyze the properties of the cosmology using the {\it Dynamical System Approach} (DSA). We will see that in spite of its original complexity ${\mathcal R}^n$- Gravity with Torsion can be easily analyzed with this method and we will be able to highlight some interesting effect that torsion and spin generate in these cosmologies. Our analysis will start with the spinless case and successively we will treat the complete problem.

%%%%%%%%%%%%%%%%%%%%%%%%%%%%%%%%%%%%%%%%%%%%%%%%%%
\subsection{The case of zero Spin}
%%%%%%%%%%%%%%%%%%%%%%%%%%%%%%%%%%%%%%%%%%%%%%%%%%
In the case of absence of spin  the comological  equations (\ref{RayRnS}-\ref{SpinEqRn}) become
\begin{eqnarray}
&& \fl \dot{\Theta}=-\frac{1 }{3}\Theta ^2-\frac{ (n-2)^{-\frac{1}{n}} \alpha ^{-\frac{1}{n}}(-2 n+3 w+3) (3w-1)^{\frac{1}{n}}}{(3 w-1) (3 (n-1) w+n-3)}\mu ^{\frac{1}{n}} 
   + \frac{4n\, \sigma ^2}{(3 (n-1) w+n-3)} ,\label{RayRn}\\
&& \fl  \dot{\sigma}=\frac{ n w-w-1}{n}\Theta  \sigma,\label{ShearRn}\\
&& \fl  2 \sigma ^2-\frac{ [3 (n-1)w+n-3]^2}{6 n^2}\Theta ^2+\frac{ [3 (n-1) w+n-3] (3 w-1)^{\frac{1}{n}}}{n\,\alpha ^{\frac{1}{n}} (n-2)^{\frac{1}{n}} (3 w-1)}\mu^{\frac{1}{n}}=0,\label{HamiCosntr}\\
&& \fl \dot{\mu}^m+\Th\mu^m(1+w) =0,\label{MatCons}
\end{eqnarray}
In order to use the DSA, let us define the dimensionless variables 
\begin{equation} \label{Var}
\fl \Omega^2=\frac{6 (n-2)^{-\frac{1}{n}} n \alpha ^{-\frac{1}{n}} \mu ^{\frac{1}{n}} (3
   w-1)^{\frac{1}{n}-1}}{\Theta ^2 (3 (n-1) w+n-3)}, \qquad \Sigma^2=\frac{12 n^2
   \sigma^2}{\Theta ^2 (3 (n-1) w+n-3)^2}.
\end{equation}
Note that the first of these variables is defined for
\begin{equation} \label{}
\frac{n}{(3 w-1) (3 (n-1) w+n-3)}>0,\qquad\frac{3 w-1}{\alpha (n-2)}>0,
\end{equation}
if $n$ is even or a rational number with even numerator and odd denominator; for
\begin{equation} \label{}
\frac{n}{\alpha  (n-2) (3 (n-1) w+n-3)}>0,
\end{equation}
if $n$ is a rational number with odd numerator and even denominator; for
\begin{equation} \label{}
\frac{n}{(3 w-1) (3 (n-1) w+n-3)}>0,
\end{equation}
if $n$ is odd or a rational number with odd numerator and odd denominator. When the above inequalities are not satisfied one has to redefine the first of \rf{Var} with a minus sign. In the following we will give  a unified treatment when possible.

Using the logarithmic time ${\mathcal N}=\ln a$, the system above can be rewritten as
\begin{eqnarray}
&& \frac{d\Omega({\mathcal N})}{d{\mathcal N}}=\Omega \left\{\frac{2 n-3 (w+1)}{6 n}\left[1\mp\Omega^2\right] + \frac{3 (n-1) w+n-3}{3n} \Sigma ^2\right\},\label{DynSys1}\\
&&\frac{d\Sigma({\mathcal N})}{d{\mathcal N}}= \Sigma \left\{\frac{3 (n-1) w+n-3}{3n}[1-\Sigma ^2]\mp\frac{2 n-3 (w+1)}{6 n}\Omega^2\right\}\label{DynSys2}\\
&& 1\mp \Omega^2-\Sigma^2=0,\label{constraint}
\end{eqnarray}
where the choice of the signs depends upon the satisfaction of the inequalities by the choice of $n$ and $w$. 
Looking at the structure of the above system it is clear that the phase space is symmetrical for reflection and has two invariant submanifolds. The first is forbidden by the fact that it would correspond to absence of matter and in this case the \rf{RofMuRn} is meaningless. The second represents isotropic cosmologies. Note also that for  $n=\frac{3(1+w)}{2}$  and $n=\frac{3(1+w)}{1+w}$ the equations decouple.

Using the \rf{constraint} one can reduce this system to a single equation. If we choose the one for $\Sigma$  we obtain
\begin{equation} \label{}
\frac{d\Sigma}{d{\mathcal N}}= \frac{1+w-2nw}{2 n} \left(1-\Sigma ^2\right) \Sigma.
\end{equation}
This equation  admits an exact solution which shows the behavior of the ``shear parameter'' with the scale factor $a$
\begin{equation}
\Sigma(a)=\frac{1}{\sqrt{e^{2 c_1} \left(\frac{a}{a_0}\right)^{\frac{w+1}{n}-2 w}+1}},
\end{equation}
where  $C$ and $S_0$ are constant. The same can be done with the other parameter obtaining
\begin{equation}
\frac{d\Omega}{d{\mathcal N}}=\mp \frac{1+w-2nw}{2 n} \left(1-\Omega ^2\right) \Omega,
\end{equation}
which admits the solution in terms of the scale factor
\begin{equation}
\Omega(a)=\mp\frac{1}{\sqrt{e^{2 c_1} \left(\frac{a}{a_0}\right)^{2 w-\frac{w+1}{n}}+1}}.
\end{equation}
It is easy to prove by direct substitution that the above solutions satisfy all of the (\ref{DynSys1}-\ref{constraint}).

In spite of these results the full dynamics of the cosmology still needs to be characterized by a phase space analysis. In the present case the phase space is one dimensional and there are two fixed points which are illustrated in Table \ref{TablePF}. 

The solutions associated with the fixed point can be found integrating the equations
\begin{eqnarray}
&&\frac{1}{\Th}\frac{d\Th({\mathcal N})}{d{\mathcal N}}=-\frac{1+w- n w}{ n}\Sigma^2, \\
&&\frac{1}{\sig}\frac{d\sig({\mathcal N})}{d{\mathcal N}} =\frac{(n-1) w-1}{n},\\
&&\frac{1}{\mu}\frac{d\mu({\mathcal N})}{d{\mathcal N}} =(1+w),
\end{eqnarray}
which are also shown in Table \ref{TablePF}. We see, then, that $\mathcal{A}$ represents the shear dominated era and $\mathcal{B}$  represents an isotropic cosmology.

Looking at the stability of these two points, $\mathcal{A}$  and $\mathcal{B}$  can be either attractors or repellers, but they never have the same character. Therefore our result shows that there are values of $n$ for which this type of cosmology tends to dissipate anisotropies and to reduce to an isotropic state.  Such values are given in Table \ref{TablePFStab}. 

%%%%%%%%%%%%%%%%%%%%%%%%%%%%%%%%%%%%%%%%%%%%%%%%%%
\begin{table}[tbp] \centering
\caption{The fixed points and eigenvalues for $R^n$-gravity with torsion in a LRS
Bianchi I model with spin.}
\begin{tabular}{cccccccc}
& & \\
\hline   Point &$(\Omega,\Sigma)$ &  Scale Factor & Energy Density& Shear \\ \hline
& & \\
$\mathcal{A}$ &$\left(0,1\right)$&  $a=a_0\left(t-t_0\right)^{\frac{n}{3[1-(n-1)w]}}$ &$\mu=0$& $\sigma=\sig_0 (t-t_0)^{-1}$\\ 
& & \\
$\mathcal{B}$& $\left(1,0\right)$& $a=a_0\left(t-t_0\right)^{\frac{2n}{3(1+w)}}$ &$\mu=\mu_0 (t-t_0)^{-2n}$& $\sigma=0$ \\
& & \\ \hline
\end{tabular}\label{TablePF}
\end{table}
%%%%%%%%%%%%%%%%%%%%%%%%%%%%%%%%%%%%%%%%%%%%%%%%%%

%%%%%%%%%%%%%%%%%%%%%%%%%%%%%%%%%%%%%%%%%%%%%%%%%%
\begin{table}[tbp] \centering
\caption{The fixed points and eigenvalues for $R^n$-gravity with torsion in a LRS
Bianchi I model.}\label{TablePFStab}
\begin{tabular}{cccccccc}
& & \\
\hline   Point &$(\Omega,\Sigma)$ & Repeller ($w=0$)& Attractor($w=0$)  \\ \hline
& & \\
$\mathcal{A}$ &$\left(0,1\right)$ &$ n<0$ &$ n>0$\\ 
& & \\
$\mathcal{B}$& $\left(1,0\right)$&$n>0$ & $n<0$ \\
& & \\ \hline
\end{tabular}
\begin{tabular}{cccccccc}
& & \\
\hline   Point &$(\Omega,\Sigma)$ &  Repeller ($0<w\leq 1$)& Attractor($0<w\leq 1$) \\ \hline
& & \\
$\mathcal{A}$ &$\left(0,1\right)$ & $n<0\lor n>\frac{1+w}{2 w}$ &$0<n<\frac{1+w}{2 w}$\\ 
& & \\
$\mathcal{B}$& $\left(1,0\right)$& $0<n<\frac{1+w}{2 w}$ &$n<0\lor n>\frac{1+w}{2 w}$ \\
& & \\ \hline
\end{tabular}
\end{table}
%%%%%%%%%%%%%%%%%%%%%%%%%%%%%%%%%%%%%%%%%%%%%%%%%%

%%%%%%%%%%%%%%%%%%%%%%%%%%%%%%%%%%%%%%%%%%%%%%%%%%
\subsection{Introducing the Spin.}
%%%%%%%%%%%%%%%%%%%%%%%%%%%%%%%%%%%%%%%%%%%%%%%%%%
We now introduce the spin of matter considering the full equations (\ref{RayRnS}-\ref{SpinEqRn}). We could perform the analysis using the variables of the previous section, but it is more convenient to redefine them as
\begin{equation} \label{Var2}
\fl \Omega=\frac{6 (n-2)^{-\frac{1}{n}} n \alpha ^{-\frac{1}{n}} \mu ^{\frac{1}{n}} (3
   w-1)^{\frac{1}{n}-1}}{\Theta ^2 (3 (n-1) w+n-3)}, \qquad \Sigma=\frac{12 n^2
   \sigma^2}{\Theta ^2 (3 (n-1) w+n-3)^2},
\end{equation}
and naturally add  the additional variable:
\begin{equation} \label{}
\Xi=\frac{4^{n-1} 3^{2n-1} n^{ 2(n-1)} }{(3 w-1)^{2(n-1)} [3 (n-1) w+n-3]^{2n}}\frac{\mathcal S^2 }{\alpha^2 \Theta ^{2(2 n-1)} },
\end{equation}
which accounts for the degrees of freedom of the spin.

In this way the equations (\ref{RayRnS}-\ref{SpinEqRn}) can be written as the dynamical system
\begin{eqnarray}
&& \fl \frac{d\Omega({\mathcal N})}{d{\mathcal N}}=\Omega \left\{\frac{2 n-3 (w+1)}{6 n}\left[1-\Omega\right] + \frac{3 (n-1) w+n-3}{3n} \Sigma 
\right.\\&&\left.~~~~~~~~~~~~ \fl+\frac{2 (3 (n-1) w+n-3) }{3 n}\Omega ^{2 (1-n)}\Xi\right\},\\
&&\fl \frac{d\Sigma({\mathcal N})}{d{\mathcal N}}= \Sigma \left\{\frac{3 (n-1) w+n-3}{3n}\left[1-\Sigma+\Omega ^{2 (1-n)}\Xi\right]-\frac{2 n-3 (w+1)}{6 n}\Omega\right\},\\
&& \fl \nn\frac{d\Xi({\mathcal N})}{d{\mathcal N}}=\Xi  \left\{\frac{4 (n-2)}{3}+\frac{2 (2 n-1)  [3 (n-1) w+n-3] }{3 n}\Xi \Omega ^{2-2 n} \right.
\\&&  ~~~~~~~~~~~~~\fl \left.-\frac{(2 n-1) [2 n-3 (w+1)]}{3 n} \Omega -\frac{2 (2 n-1) \Sigma  (3 (n-1) w+n-3)}{3 n}\right\},\\
&& \fl 1-\Sigma- \Omega+\Xi \Omega^{2(1-n)}=0.\label{constraintS}
\end{eqnarray}
Using the \rf{constraintS},  the equation for $\Xi$  can be eliminated
\begin{eqnarray}
&& \frac{d\Omega({\mathcal N})}{d{\mathcal N}}=\frac{ 1+w-2 n w}{n}\Omega \left(1-\Omega\right),\label{DynSysTS1}\\
&&\frac{d\Sigma({\mathcal N})}{d{\mathcal N}}=-\frac{1+w-2 nw}{n}\Sigma\,  \Omega,\label{DynSysTS2}\\
&& 1- \Omega+\Xi \Omega^{2(1-n)}-\Sigma=0,
\end{eqnarray}
and it is clear that the equation for $\Xi$ is decoupled and can be neglected. This system possesses two invariant submanifolds $\Sigma=0$ and $\Omega=0$.  For $n\neq0$ and $1+w-2 n w=0$ the system becomes decoupled (and trivial) and can be solved exactly to give $\mu\propto \Theta^{\frac{1+w}{w}}$,  $\Sigma\propto \Theta^{-\frac{(n-1)w-1}{n }}$,  $\mathcal{S}\propto \Theta^{-1}$ and 
\begin{equation}
\dot{\Theta}=-w\Theta^2,
\end{equation}
which means
\begin{eqnarray}
&&a=a_0 (t-t_0)^{1/w},\\
&&\mu=\mu_0 (t-t_0)^{-\frac{1+w}{w}},\\
&&\sigma=\sigma_0 (t-t_0)^{-\frac{(n-1)w-1}{n w}},\\
&& \mathcal{S}=\mathcal{S}_0 (t-t_0)^{-1/w}
\end{eqnarray}
Using  also the equation for $\Xi$ in which \rf{constraintS} is implemented
\begin{equation}
\frac{d\Xi({\mathcal N})}{d{\mathcal N}}=-\frac{1+w-2 n w}{n} [(2 n-1) \Omega -2 n+2]\Xi
\end{equation}
it is easy to check that the above solution satisfies all the equations.

Setting the L.H.S of the (\ref{DynSysTS1}-\ref{DynSysTS2}) to zero we find the fixed points in Table \ref{SolFP}. The system presents a line of fixed points and one single fixed point. The stability of these fixed points can be found using the  Hartman-Grobman (HG) theorem and it is given together with their coordinates in Table \ref{TableStab}. 

The solutions for the scale factor associated to the fixed points can be found solving the equations 
\begin{eqnarray}
&&\frac{1}{\Th}\frac{d\Th({\mathcal N})}{d{\mathcal N}}=\frac{(n-1) w-1}{n}+\frac{1+w-2 n w}{2 n}\Omega,\\
&&\frac{1}{\sig}\frac{d\sig({\mathcal N})}{d{\mathcal N}} =\frac{(n-1) w-1}{n},\\
&&\frac{1}{\mu}\frac{d\mu({\mathcal N})}{d{\mathcal N}} =(1+w),\\
&&\frac{1}{\mathcal S}\frac{d{\mathcal S}({\mathcal N})}{d{\mathcal N}} =-1.
\end{eqnarray}
The complete solutions are shown in Table \ref{SolFP}. 

%%%%%%%%%%%%%%%%%%%%%%%%%%%%%%%%%%%%%%%%%%%%%%%%% 
\begin{table}[tbp] \centering
\caption{The solutions of the scale factor and shear evolution for $R^n$-gravity in a LRS Bianchi I model.}\label{SolFP}
\begin{tabular}{cclllll}
& &  \\
\hline  Fixed Manifolds &  ($\Omega, \Sigma, \Xi$)&Scale factor & Shear\\ \hline
& & \\
 $\mathcal{L}$ & (0,$\Sigma_0$,0) &$a=a_0\left(t-t_0\right)^{\frac{n}{3[1-(n-1)w]}}$ & $\sigma=\sig_0 (t-t_0)^{-1}$ \\
  $\mathcal{A}$ & (1,0,0) &$a=a_0\left(t-t_0\right)^{\frac{2n}{3(1+w)}}$ & $\sigma=
  0$& \\
& & \\ \hline
\end{tabular}
\begin{tabular}{cclllll}
& &  \\
\hline  Fixed Manifolds &  ($\Omega, \Sigma, \Xi$)& Energy Density& Spin Scalar
\\ \hline
& & \\
 $\mathcal{L}$ & (0,$\Sigma_0$,0) &$\mu=0$& ${\mathcal S}={\mathcal S}_0(t-t_0)^{-\frac{n}{[1-(n-1)w]}}$ \\
  $\mathcal{A}$ & (1,0,0) &$\mu=\mu_0 (t-t_0)^{-2n}$&${\mathcal S}= 0$ \\
& & \\ \hline
\end{tabular}
\end{table}
%%%%%%%%%%%%%%%%%%%%%%%%%%%%%%%%%%%%%%%%%%%%%%%%%%%
%%%%%%%%%%%%%%%%%%%%%%%%%%%%%%%%%%%%%%%%%%%%%%%%%%
\begin{table}[tbp] \centering
\caption{The stability of the fixed points for $R^n$-gravity with torsion in a LRS
Bianchi I cosmology with spin. Here we always assume $w\neq1/3$.  }
\begin{tabular}{ccccccccc}
\hline Point & Eigenvalues && Attractor& Repeller & \\
$\mathcal{L}$ &$\left\{0, \frac{1+w-2 n w}{2 n}\right\}$&
&$n>0\land  w=0$ &$n<0\land  w=0$\\
& &&$0<w\leq 1\land \left(n<0\lor n>\frac{w+1}{2 w}\right)$&$0<w\leq 1\land 0<n<\frac{w+1}{2 w}$ \\
& & \\
$\mathcal{B}$ &$\left\{-\frac{1+w-2 n w}{2 n},-\frac{1+w-2 n w}{2 n}\right\}$&&$0<n<1\land 0\leq w\leq 1$  &$n<0\land 0\leq w\leq 1$\\
& && $n\geq 1\land 0\leq w<\frac{1}{2 n-1}$&$n>1\land \frac{1}{2 n-1}<w\leq 1$  \\
& & \\ \hline
\end{tabular}\label{TableStab}
\end{table}
%%%%%%%%%%%%%%%%%%%%%%%%%%%%%%%%%%%%%%%%%%%%%%%%%%

\begin{figure}[htbp]
\includegraphics[scale=0.9]{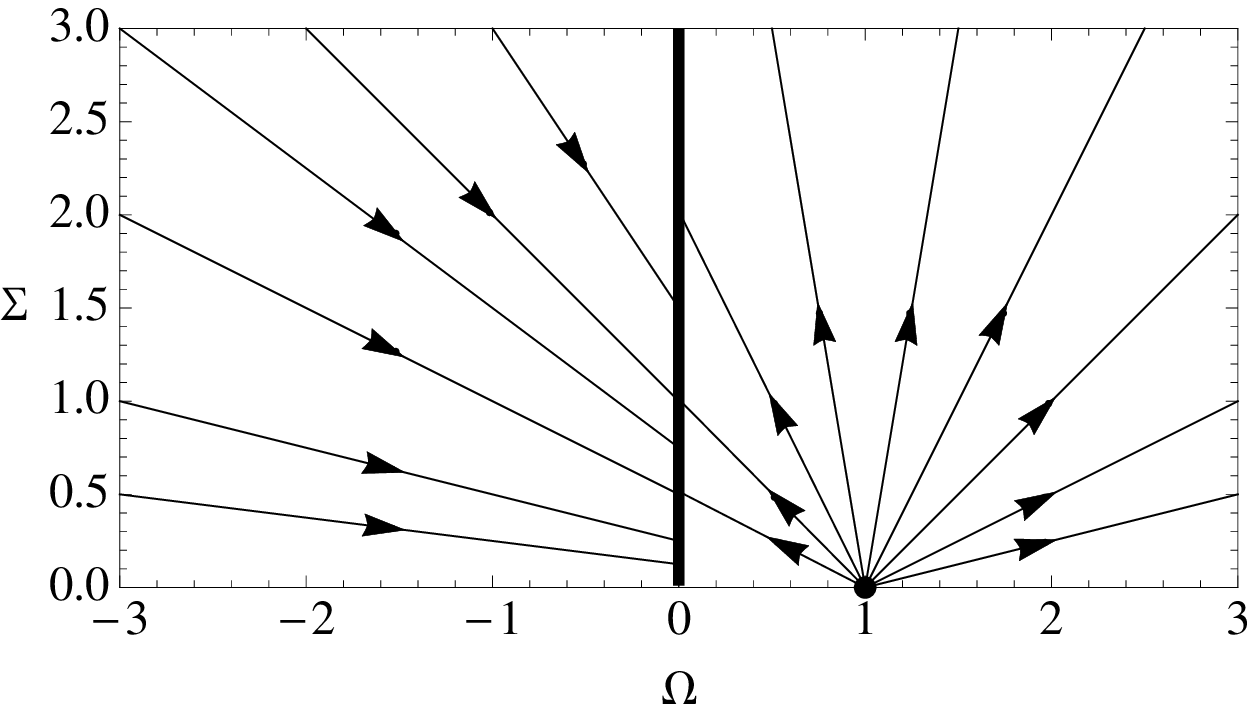}
\centering
\caption{Phase space of the  $f(R)$ gravity with torsion in Bianchi I spacetime  for $n<0$, $w=0$. The matter variable can be negative for special values of the parameters (see \rf{Var2}).  }
\label{PS1}
\end{figure}

\begin{figure}[htbp]
\includegraphics[scale=0.9]{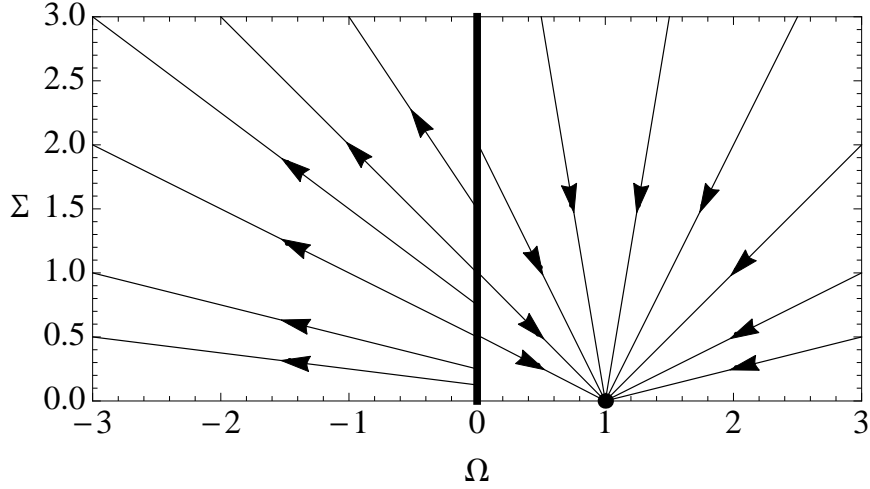}
\centering
\caption{Phase space of the  $f(R)$ gravity with torsion in Bianchi I spacetime  for $n>1$, $w=0$. The matter variable can be negative for special values of the parameters (see \rf{Var2}).}
\label{PS2}
\end{figure}

Plots of the phase space in the case of dust  ($w=0$) cosmologies are given in Figures \ref{PS1}-\ref{PS2}. 

%%%%%%%%%%%%%%%%%%%%%%%%%%%%%%%%%%%%%%%%%%%%%%%%%
%%%%%%%%%%%%%%%%%%%%%%%%%%%%%%%%%%%%%%%%%%%%%%%%%
%%%%%%%%%%%%%%%%%%%%%%%%%%%%%%%%%%%%%%%%%%%%%%%%%
%%%%%%%%%%%%%%%%%%%%%%%%%%%%%%%%%%%%%%%%%%%%%%%%%

%%%%%%%%%%%%%%%%%%%%%%%%%%%%%%%%%%%%%%%%%%%%%%%%%%
\subsubsection{Asymptotics.}\label{Asy}
%%%%%%%%%%%%%%%%%%%%%%%%%%%%%%%%%%%%%%%%%%%%%%%%%%
The system (\ref{DynSys1}-\ref{constraint}) is not compact therefore it is necessary to examine the asymptotic regime. 
This  is done normally using the Poincar\`{e} projection \cite{Lefschetz,Carloni:2004kp},  however the structure of (\ref{DynSys1}-\ref{constraint})
allows us to use an alternative method, suggested for the first time in \cite{Goliath:1998na}.

Starting from the \rf{HamiCosntr} we define the variable 
\begin{equation} \label{}
\fl\mathfrak{D}=\sqrt{\Theta ^2+\frac{3 (n-2)^{2-\frac{2}{n}}\alpha ^{-2/n} (3 w-1)^{\frac{2}{n}-2} }{ [3 (n-1)w+n-3]^2} {\mathcal S}^2\,\mu^{\frac{2}{n}-2}}=|\Th|\sqrt{1+\Xi^2 \Omega^{4(1-n)}},
\end{equation}
and 
\begin{eqnarray}\label{VarAsy}
&&\tilde{\Sigma}=2 \sqrt{3}\left|\frac{ n}{ (3 (n-1)w+n-3)}\right|\frac{\sigma}{\mathfrak{D}},\qquad Q=\frac{\Theta }{\mathfrak{D}},\label{VarAsy1}\\
&&\tilde{\Omega}^2=\pm\frac{6 (n-2)^{-\frac{1}{n}} n \alpha ^{-\frac{1}{n}} (3 w-1)^{\frac{1}{n}-1}}{(3 (n-1) w+n-3)}\frac{ \mu^{\frac{1}{n}}}{ \mathfrak{D}}\label{VarAsy2}.
\end{eqnarray}
where the $\pm$ ensures a meaningful definition of $\tilde{\Omega}$ for every choice of $n$ and $w$. Defining the time coordinate $\tau$ such that $X_{,\tau}=\mathfrak{D}^{-1}\dot{X}$  the  (\ref{DynSys1}-\ref{constraint}) can be written as
\begin{eqnarray}
&&\frac{d\mathfrak{D} (\tau)}{d\tau}=\mathfrak{D} Q \left[\frac{3 w (n-1) +n-3}{3 n}\tilde{\Sigma}^2\mp\frac{ 3 (w+1)-2n}{6 n}\tilde{\Omega}^2-\frac{1}{3}\right],\\
&& \frac{dQ(\tau)}{d\tau}=\frac{3 w(n-1) +n-3}{3 n}\left(\tilde{\Sigma}^2-1\right)\mp\frac{3 (w+1)-2 n}{6 n}\tilde{\Omega}^2 (1+Q^2),\\
&&\frac{d\tilde{\Omega}^2(\tau)}{d\tau}=\frac{1+w- 2 n w}{2 n}Q\,\tilde{\Omega}^2\left( 1\pm\tilde{\Omega}^2 \right),\\
&&\frac{d\tilde{\Sigma}(\tau)}{d\tau}=\mp\frac{ (2 n w-w-1)}{2 n}Q \,\tilde{\Omega}^2\,\tilde{\Sigma}^2,\\
&&\tilde{\Sigma}^2 \pm\tilde{\Omega}^2=1.\label{costraintB}
\end{eqnarray}
where the $\pm$ and $\mp$ are related to the definition of $\tilde{\Omega}$. It is clear from the previous system and the definitions (\ref{VarAsy1}-\ref{VarAsy2}) that if the $\tilde{\Omega}$ is defined with a ``+'' sign the system  above is compact  ($0<Q<1$, $0<\tilde{\Omega}<1$).

Using the constraint one can eliminate the equation for $\tilde{\Sigma}$
\begin{eqnarray}
&&\frac{d\mathfrak{D} (\tau)}{d\tau}=\left(\frac{(n-1) w-1}{n}\pm\frac{ 1+w-2 n w}{2 n}\tilde{\Omega}^2\right)Q\,\mathfrak{D},\label{asy1}\\
&& \frac{dQ(\tau)}{d\tau}=\pm\frac{ (1+w-2 n w)}{2 n}\Omega^2(1-Q^2),\\
&&\frac{d\tilde{\Omega}(\tau)}{d\tau}=\frac{1+w- 2 n w}{2 n}Q\,\tilde{\Omega}\left( 1\mp\tilde{\Omega}^2 \right),\\
&&\tilde{\Sigma^2} \pm\tilde{\Omega}^2=1\label{asy4}.
\end{eqnarray}
It is clear that the first equation is decoupled so the dynamical system is only composed by the last three equations. The fixed manifolds can be found in the usual way and are indicated in Table \ref{SolFPB}. One of the two fixed points  ($\tilde{\mathcal{A}}_+$) have the same character of  $\mathcal{A}$ and the  other  ($\tilde{\mathcal{A}}_-$) corresponds to the contracting versions of the solutions of $\mathcal{A}$. In addition to $\tilde{\mathcal{A}}_\pm$ we find an entire fixed line  ($\tilde{\mathcal{L}}$) which contains also a fixed point representing  a power law solution that depends on the parameters $n$ and $w$. The line also contains the origin point which corresponds to a Static Einstein universe.

%%%%%%%%%%%%%%%%%%%%%%%%%%%%%%%%%%%%%%%%%%%%%%%%% 
\begin{table}[tbp] \centering
\caption{The fixed points of the asymptotic analysis of Section \ref{Asy}}
\begin{tabular}{cclllll}
& &  \\
\hline  Point &  ($Q,\tilde{\Omega}$)&Scale factor & Shear& Energy Density& Spin Scalar
\\ \hline
& & \\
 $\tilde{\mathcal{A}}_-$ & $(-1,1)$ &$a=a_0\left(t-t_0\right)^{-\frac{2n}{3(1+w)}}$ & $\sigma=0$&$\mu=\mu_0 (t-t_0)^{-2n}$& ${\mathcal S}=0$ \\
  $\tilde{\mathcal{A}}_+$ & $(1,1) $&$a=a_0\left(t-t_0\right)^{+\frac{2n}{3(1+w)}}$ & $\sigma=0$&$\mu=\mu_0 (t-t_0)^{2n}$\\
   $\tilde{\mathcal{L}}$ &$\left(Q_0,0\right)$&$a=a_0\left(t-t_0\right)^{\frac{\sgn(Q_0) n}{3[1-(n-1)w]}}$ & $\sigma=\sig_0 (t-t_0)^{-\sgn(Q_0)}$&$\mu=0$& $ {\mathcal S}=0$ \\
& & \\ \hline
\end{tabular}\label{SolFPB}
\end{table}
%%%%%%%%%%%%%%%%%%%%%%%%%%%%%%%%%%%%%%%%%%%%%%%%%%%

The stability of the fixed manifolds can be also calculated with the HG theorem and it is shown in Table \rf{TableStabB}.  For the points common to the ones of the previous section the results are, naturally, consistent. The fixed line instead has a stability that depends on the sign of $\frac{ Q_0(1 + w - 2 n w))}{(2 n)}$ where $Q_0$ represents the point on the line. The point (0,0) has two zero eigenvalues and is not isolated. This tells us that, on the one hand, the point is unstable and on the other that the orientation of the orbits close to this point depends on the one of the neighboring points.

%%%%%%%%%%%%%%%%%%%%%%%%%%%%%%%%%%%%%%%%%%%%%%%%%%
\begin{table}[tbp] \centering
\caption{The stability of the fixed points for $R^n$-gravity with torsion in a LRS
Bianchi I cosmology with spin in the asymptotic analysis. Here we always assume $w\neq1/3$. Note that the stability of the ``+'' points is the same of the one in the standard analysis and the one of the ``-'' points is reversed, as expected.}
\begin{tabular}{ccccccccc}
\hline Point & Eigenvalues && Attractor& Repeller \\ \hline\\
$\tilde{\mathcal{A}}_-$ &$\left\{\frac{-2 n w+w+1}{n},\frac{-2 n w+w+1}{n}\right\}$&
& $n < 0\land 0\leq w\leq 1$ & $0 < n < 1\land 0\leq w\leq 1$ \\
& && $n > 1\land\frac {1} {2 n - 1} <  w\leq 1$ & $n\geq 1\land 0\leq w < \frac {1} {2 n - 1}$ \\ \\
$\tilde{\mathcal{A}}_+$ &$\left\{-\frac{-2 n w+w+1}{n},-\frac{-2 n w+w+1}{n}\right\}$&& $0 < n < 1\land 0\leq w\leq 1$ & $n < 0\land 0\leq w\leq 1$\\
& && $n\geq 1\land 0\leq w < \frac {1} {2 n - 1} $ & $n>1\land \frac{1}{2 n-1}<w\leq 1$\\ \\
$\tilde{\mathcal{L}}$ &$\left\{\frac{(1+w-2 n w)Q_0}{n},0\right\}$&
$Q_0>0$& $n < 0\land 0\leq w\leq 1$ & $0 < n < 1\land 0\leq w\leq1 $ \\
& && $n > 1\land\frac {1} {2 n - 1} <  w\leq 1$ & $n\geq 1\land 0\leq w < \frac {1} {2 n - 1}$ \\&&
$Q_0<0$& $0 < n < 1\land 0\leq w\leq1 $& $n < 0\land 0\leq w\leq 1$  \\
& && $n\geq 1\land 0\leq w < \frac {1} {2 n - 1}$& $n > 1\land\frac {1} {2 n - 1} <  w\leq 1$  \\ 
& &&  \\ \hline
\end{tabular}\label{TableStabB}
\end{table}
%%%%%%%%%%%%%%%%%%%%%%%%%%%%%%%%%%%%%%%%%%%%%%%%%%

\begin{figure}[htbp]
\includegraphics[scale=0.9]{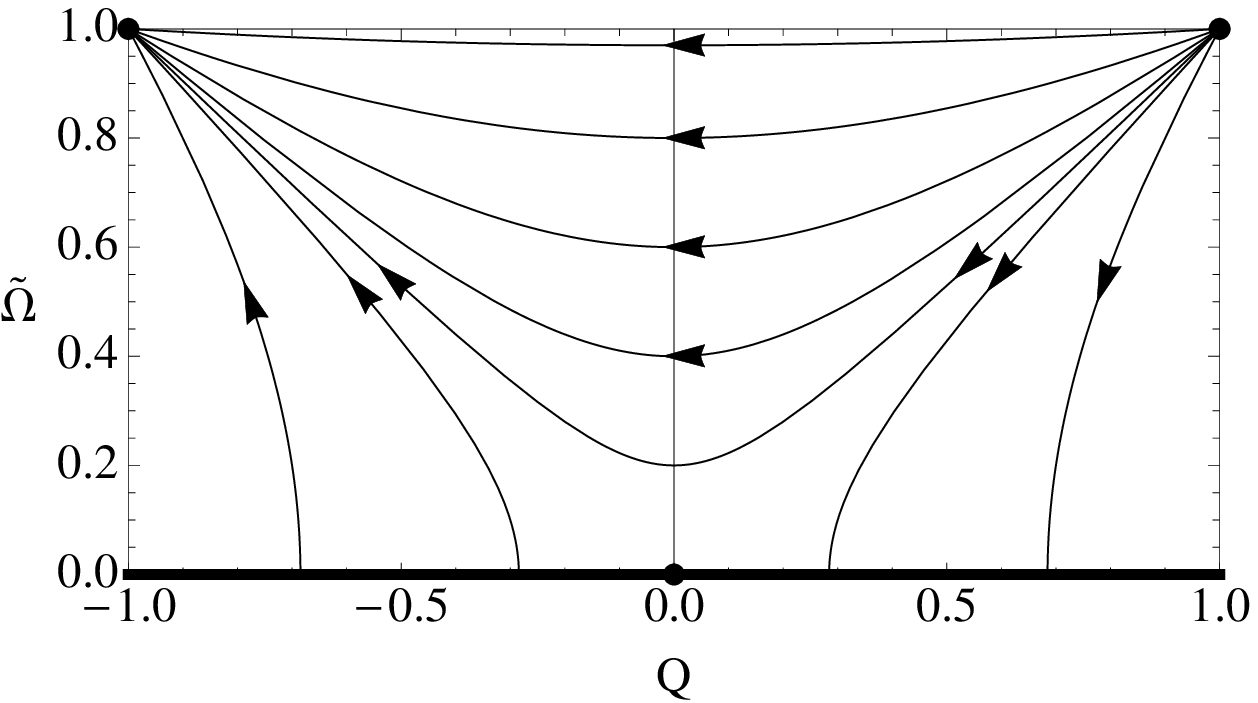}
\centering
\caption{Phase space of the system (\ref{asy1}-\ref{asy4})  for $n<0$, $w=0$ and $\tilde{\Omega}$  defined with ``+''.}
\label{fig:bounce1}
\end{figure}

\begin{figure}[htbp]
\includegraphics[scale=0.9]{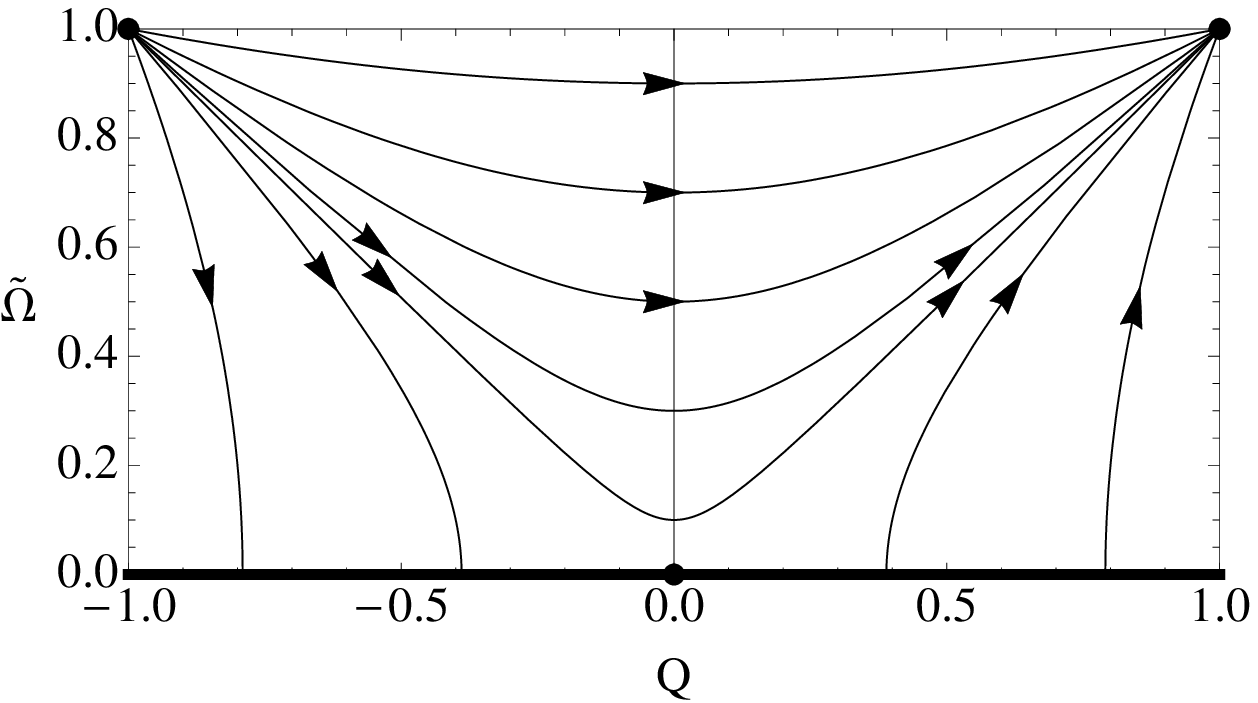}
\centering
\caption{Phase space of the system (\ref{asy1}-\ref{asy4}) for $n>1$, $w=0$ and $\tilde{\Omega}$  defined with ``+''.}
\label{fig:bounce2}
\end{figure}

\begin{figure}[htbp]
\includegraphics[scale=0.9]{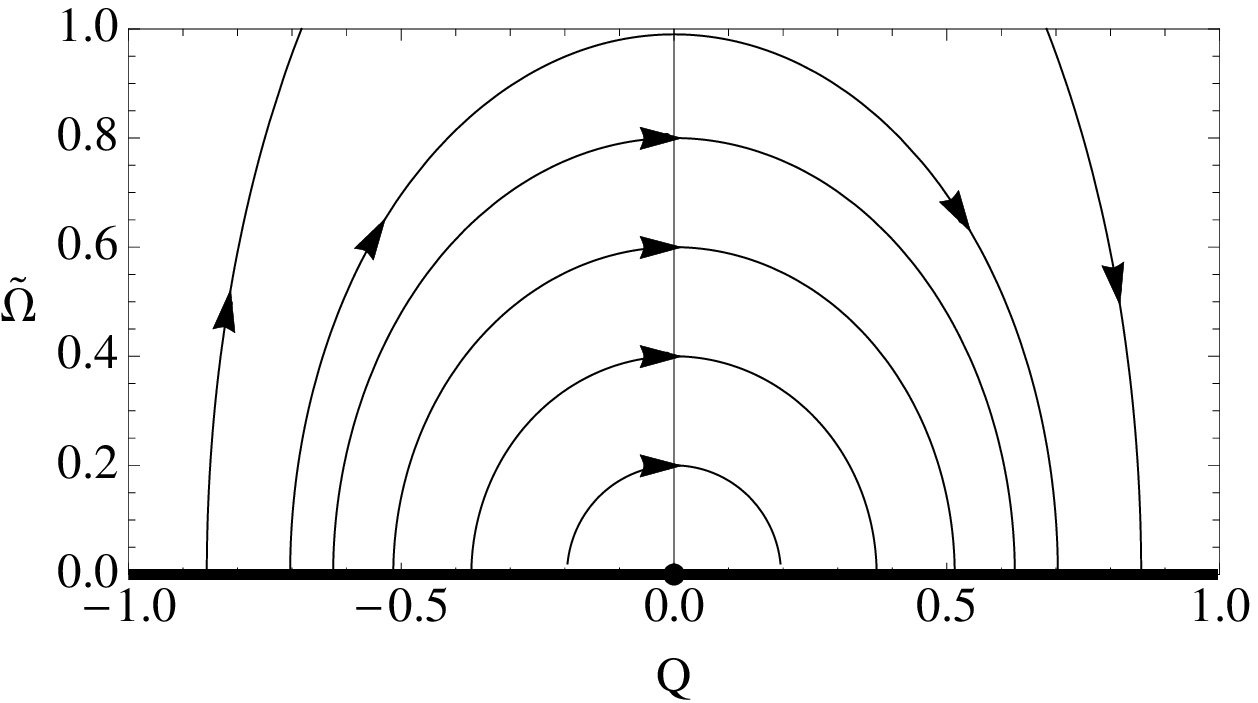}
\centering
\caption{Phase space of the system (\ref{asy1}-\ref{asy4}) for $n<0$, $w=0$ and $\tilde{\Omega}$  defined with ``-''.}
\label{fig:bounce3}
\end{figure}

\begin{figure}[htbp]
\includegraphics[scale=0.9]{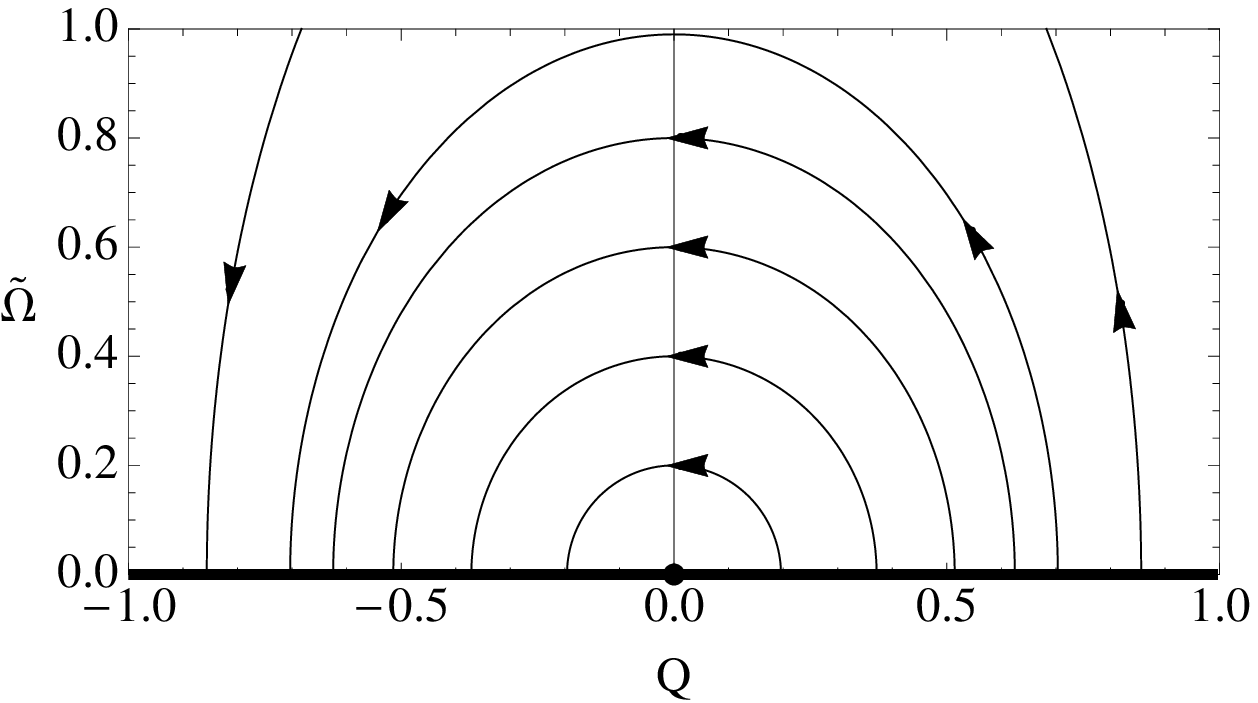}
\centering
\caption{Phase space of the system (\ref{asy1}-\ref{asy4})  for $n>1$, $w=0$ and $\tilde{\Omega}$  defined with ``-''.}
\label{fig:bounce4}
\end{figure}

Examples of the phase space are given in Figure (\ref{fig:bounce1}-\ref{fig:bounce4}).
%%%%%%%%%%%%%%%%%%%%%%%%%%%%%%%%%%%%%%%%%%%%%%%%%
%%%%%%%%%%%%%%%%%%%%%%%%%%%%%%%%%%%%%%%%%%%%%%%%%
%%%%%%%%%%%%%%%%%%%%%%%%%%%%%%%%%%%%%%%%%%%%%%%%%
%%%%%%%%%%%%%%%%%%%%%%%%%%%%%%%%%%%%%%%%%%%%%%%%%
%%%%%%%%%%%%%%%%%%%%%%%%%%%%%%%%%%%%%%%%%%%%%%%%%
%%%%%%%%%%%%%%%%%%%%%%%%%%%%%%%%%%%%%%%%%%%%%%%%%
%%%%%%%%%%%%%%%%%%%%%%%%%%%%%%%%%%%%%%%%%%%%%%%%%
%%%%%%%%%%%%%%%%%%%%%%%%%%%%%%%%%%%%%%%%%%%%%%%%%

The presence of a static solution and the fact that this solution is unstable is a sign that a bouncing solution is possible in this 
type of cosmology.  A bounce in General Relativity is characterized by the condition 
\begin{equation}\label{bounce}
\Th_0=0, \qquad \dot{\Th}_0>0.
\end{equation}
holding at the bounce instant $t_0$. However in non standard gravity, since the scale factor does not evolve monotonically  this definition is not so clearcut \cite{Carloni:2005ii}. In fact one can only characterize if the scale factor presents a minimum, but this is not in general sufficient  to realize the classical bounce scenario, which also require $|a|\ll 1$. In what follows we will focus only on the conditions \rf{bounce} i.e. the necessary conditions for a bounce,  leaving a more detailed analysis for a dedicated future work.

From our phase space analysis it is clear that  the theory presents a bouncing behaviour realized when the matter variable is chosen with the plus sign i.e. 
\begin{equation} \label{Bounce1}
\frac{6 (n-2)^{-\frac{1}{n}} n \alpha ^{-\frac{1}{n}} (3 w-1)^{\frac{1}{n}-1}}{(3 (n-1) w+n-3)}>0, \qquad\mbox{and}\qquad n>1,
\end{equation}
and describes a cosmology that, starting form an isotropic collapsing state bounces approaching an isotropic expansion state. It is interesting to note that, in a  counterintuitive way, the matter density decreases  during the collapsing phase, so that in the bounce it is lower than the starting value. This feature is probably due to the complex interaction between higher order terms and matter.
Instead, when  
\begin{equation} \label{Bounce2}
\frac{6 (n-2)^{-\frac{1}{n}} n \alpha ^{-\frac{1}{n}} (3 w-1)^{\frac{1}{n}-1}}{(3 (n-1) w+n-3)}<0,  \qquad\mbox{and}\qquad n<0,
\end{equation}
we have bounce connecting two shear dominated solutions  i.e. a cosmology that, starting form an anisotropic collapsing state approaches to an anisotropic expansion state. In this case the matter energy density grows with the collapse, and the switch between expansion and contraction happens for a value of energy density higher than the initial one.

%%%%%%%%%%%%%%%%%%%%%%%%%%%%%%%%%%%%%%%%%%%%%%%%%%
\subsection{Conclusion}
%%%%%%%%%%%%%%%%%%%%%%%%%%%%%%%%%%%%%%%%%%%%%%%%%%
In this paper we have analyzed the Bianchi I cosmology of $f(R)$-gravity with torsion using the Dynamical Systems Approach.
Compared with the standard $f(R)$-gravity, theories with torsion contain the additional constraint \rf{2.1bis} that simplifies considerably the phase space analysis. In the case of $f(\mathcal{R})=\mathcal{R}^n$ \rf{2.1bis} gives an analytical result. With this choice of $f$, we examined both the cases in which the cosmological fluid does not posses spin and the one in which the spin is present.

In the first case the phase space is unidimensional and contains two fixed points which can be attractors or repellers. In particular, for $n<0$ and $0<n<\frac{w+1}{2w}\; (\mbox{if}\; w\neq0)$, the anisotropy dominated fixed point is unstable and the matter dominated one is an attractor. This means that the cosmology tends to isotropize. This result is similar to the one found in \cite{Leach:2006br} in which the case of a Bianchi I cosmology without torsion was analyzed.  In that case, however, the limits on the values of the parameter $n$ were different, but the conclusions drawn there are valid also in this case: $f(R)$-gravity with torsion can also, provided that other observational constraints (such as the spectrum of gravitational waves) are satisfied, constitute an alternative to inflation. Indeed an isotropization effect due to torsion is already a feature of the standard EKSC theories of gravity, but it usually requires a contracting phase and a bounce \cite{Poplawski}.  In this respect the higher order corrections permit to eliminate the need for a bouncing behavior. The dynamical systems equations can be also solved exactly giving  a useful relation between the matter and shear densities which could be used to perform, for example,  the classical cosmological tests for this class of theories. 

In the case of a cosmic fluid with spin the dimension of phase space grows accordingly. The spin terms, however, do not enter directly in the  Raychaudhuri equations and therefore there is no ``spin dominated'' fixed point.  For this reason the introduction of the spin does not modify much the geometry of the orbits and, for example, the conditions on the fixed point stability for the presence of an isotropic attractor do not change. However in this case the set of initial conditions for which this scenario is realized is smaller. Specifically, the isotropization of cosmic histories can only be realized if the initial conditions are chosen in so that $0<\Omega_0<1$, which in terms of the standard cosmological parameters means
\begin{equation} \label{}
\Theta^2 \mu^{-\frac{1}{n}}>\frac{6 (n-2)^{-\frac{1}{n}} n \alpha ^{-\frac{1}{n}} (3 w-1)^{\frac{1}{n}-1}}{3 (n-1)w+n-3}>0.
\end{equation}

The  cosmologies we have analyzed also include  the necessary conditions for the occurrence of a bounce.  The classical version of the bounce i.e. a phase of isotropic collapse followed by one of isotropic expansion is found to be possible only if $n>1$  and the initial conditions associated with this specific behaviour are given by the inequality $\tilde{\Omega}_0>|Q_0|$ which, in terms of the standard cosmological quantities gives
\begin{equation} \label{}
\frac{\mu_0}{\Th_0^{2}}>\left|\frac{(n-2)^{\frac{1}{n}} \alpha ^{\frac{1}{n}} (3 w-1)^{-\frac{1-n}{n}} (3
   (n-1) w+n-3)^{\frac{1}{n}}}{6 n(3 w-1)^{\frac{1-n}{n}}}\right|.
\end{equation} 
The bouncing is also characterized by a decrease of the matter variable $\tilde{\Omega}$. Since this variable is a monotonic function of the energy density it turns out that during the contraction phase the matter density actually decreases, which might appear counterintuitive. However one must bear in mind that there is an ``effective fluid'' in the model: the one represented by the higher order corrections. Therefore such behavior is most likely due to the combination of the presence of the spin and these additional terms.

In addition, the phase space analysis also shows a more exotic form of bounce, which links a contracting and an expanding shear dominated solutions. This behavior is realized for any initial condition provided that 
\begin{equation} \label{Bounce2}
\frac{6 (n-2)^{-\frac{1}{n}} n \alpha ^{-\frac{1}{n}} (3 w-1)^{\frac{1}{n}-1}}{(3 (n-1) w+n-3)}<0  \qquad\mbox{and}\qquad n<0
\end{equation}
and the matter energy density has the expected behavior i.e. grows in the contraction phase. It would be interesting to analyze in detail this new typology of bounce and, for example, understand if the anisotropy is able to ``transmit'' information between the contraction and expansion phase.

On more general terms, our analysis confirms, consistently with the results obtained in \cite{VFC,VF2}, that  while torsion can influence in a non trivial way the behavior of $f(R)$ cosmologies, the spin has a secondary role. The most likely explanation for this difference lays in the symmetries of the spacetime and the spin tensor, which prevent a direct coupling of the two quantities. If this line of reasoning is correct one can expect  that the spin will have a key role in cosmologies in which the vorticity is non negligible, like Godel's universes, or, more interestingly, in the analysis of the gravitational collapse in which it is natural to expect an growing value of the vorticity. The exploration of these last cases will be an interesting further development of the present work.

\end{document}